\documentclass[a4paper,11pt]{article}
\usepackage{pos}

\newcommand{\ie}{{\it{i.e.}}}\newcommand{\bq}{\mbox{\boldmath $q$}}

\newcommand{\bP}{\mbox{\boldmath $P$}}
\newcommand{\bQ}{\mbox{\boldmath $Q$}}

\newcommand{\bp}{\mbox{\boldmath $p$}}
\newcommand{\bb}{\mbox{\boldmath $b$}}

\newcommand{\bE}{\mbox{\boldmath $E$}}

\title{From Wigner distributions of photons to dilepton production in semicentral heavy ion collisions}
\ShortTitle{Dilepton production in semicentral heavy ion collisions}

\author*[a]{Mariola K{\l}usek-Gawenda}
\author[a,]{Wolfgang Sch\"afer}
\author[a,b]{Antoni Szczurek}
\author[c]{Roxana Busuioc}
\author[d]{Yash Arya}

\affiliation[a]{Institute of Nuclear Physics Polish Academy of Sciences,\\ 
	ul. Radzikowskiego 152, PL-31-342 Krak\'ow, Poland}

\affiliation[b]{College of Natural Sciences, Institute of Physics, University of Rzesz\'ow, \\
	ul. Pigonia 1, PL-35-310 Rzesz\'ow, Poland}

\affiliation[c]{The University of Glasgow, \\
	University Avenue, G12 8QQ Glasgow, United Kingdom}

\affiliation[d]{Department of Physics, Hansraj College, University of Delhi, \\
	
	Mahatma Hans Raj Marg, Malka Ganj, 110007, Delhi, India}

\emailAdd{mariola.klusek@ifj.edu.pl}
\emailAdd{wolfgang.schafer@ifj.edu.pl}
\emailAdd{antoni.szczurek@ifj.edu.pl}
\emailAdd{2482934b@student.gla.ac.uk}
\emailAdd{yasharya823@gmail.com}

\abstract{We present a formalism how to calculate differential distributions for dilepton production in semicentral heavy ion collisions. In this new approach, the differential cross section is calculated using the complete polarization density matrix of photons resulting from the Wigner distribution formalism. The formalism is used to calculate different distributions of the invariant mass, dilepton transverse momentum and acoplanarity for different regions of centrality. The results of the calculation are compared with recent experimental data. We also study the contribution from different parts of impact parameter space.}

\FullConference{%
  *** The European Physical Society Conference on High Energy Physics (EPS-HEP2021), ***\\
  *** 26-30 July 2021 ***\\
  *** Online conference, jointly organized by Universität Hamburg and the research center DESY ***
}

\begin{document}
\maketitle

\section{Introduction}

Ultrarelativistic Heavy Ions of large charge $Z$ are accompanied by a large
flux of Weizs\"acker-Williams photons. This opens up the opportunity to study a
variety of photo-induced nuclear processes, as well as photon-photon processes.
However, if one interprets Weizs\"acker-Williams photons as partons of the nucleus, it appears natural that the coherent photon cloud also contributes in semicentral or central collisions, where $b<2 R_A$ where $R_A$ is a nuclear radius. In such collisions the colliding nuclei will interact strongly generating an event, which may include the production of a quark-gluon plasma (QGP).

We study the invariant mass distributions of dileptons produced in ultrarelativistic
heavy-ion collisions at very low pair transverse momenta, $P_T\leq 0.15$\,GeV, and distribution in transverse momentum of the dilepton pair for fixed centrality and invariant mass. The comparison of experimental STAR and ALICE data with theoretical results also will be discussed.  

\section{Methodology}

The coherent initial photon fusion reactions is calculated in the impact parameter space. This approach allows controling the distance between two colliding nuclei. The basic ingredient for the $\gamma\gamma$ fusion mechanism is the Weizs\"acker-Williams flux of photons $N(\omega,b)$ for an ion of charge $Z$ which moves along impact parameter $\bb$ ($b = |\bb|$) with the Lorentz-boost parameter. The differential cross section for dilepton $l^+ l^-$ production via $\gamma \gamma$ fusion at 
fixed impact parameter $\bb$ of the nucleus-nucleus collision can then be written with the phase space element $d\xi = dy_1 dy_2 dp_T^2$ \cite{Klusek-Gawenda:2018zfz}. Taking into account this form of integration, we have the possibility to put limits on the rapidity of single outgoing lepton or lepton transverse momentum, $p_T$. The presented considerations do not make it possible to take into account the distributions in the pair-$P_T$. Such a step can only be performed when we replace the variables in the photon flux. Instead of integral over $b$, we use $b$-integrated transverse momentum dependent photon fluxes (see Eq.~(\ref{eq:kt-fact})).

\begin{figure}[!h]
	\centering
	\includegraphics[width=.5\textwidth]{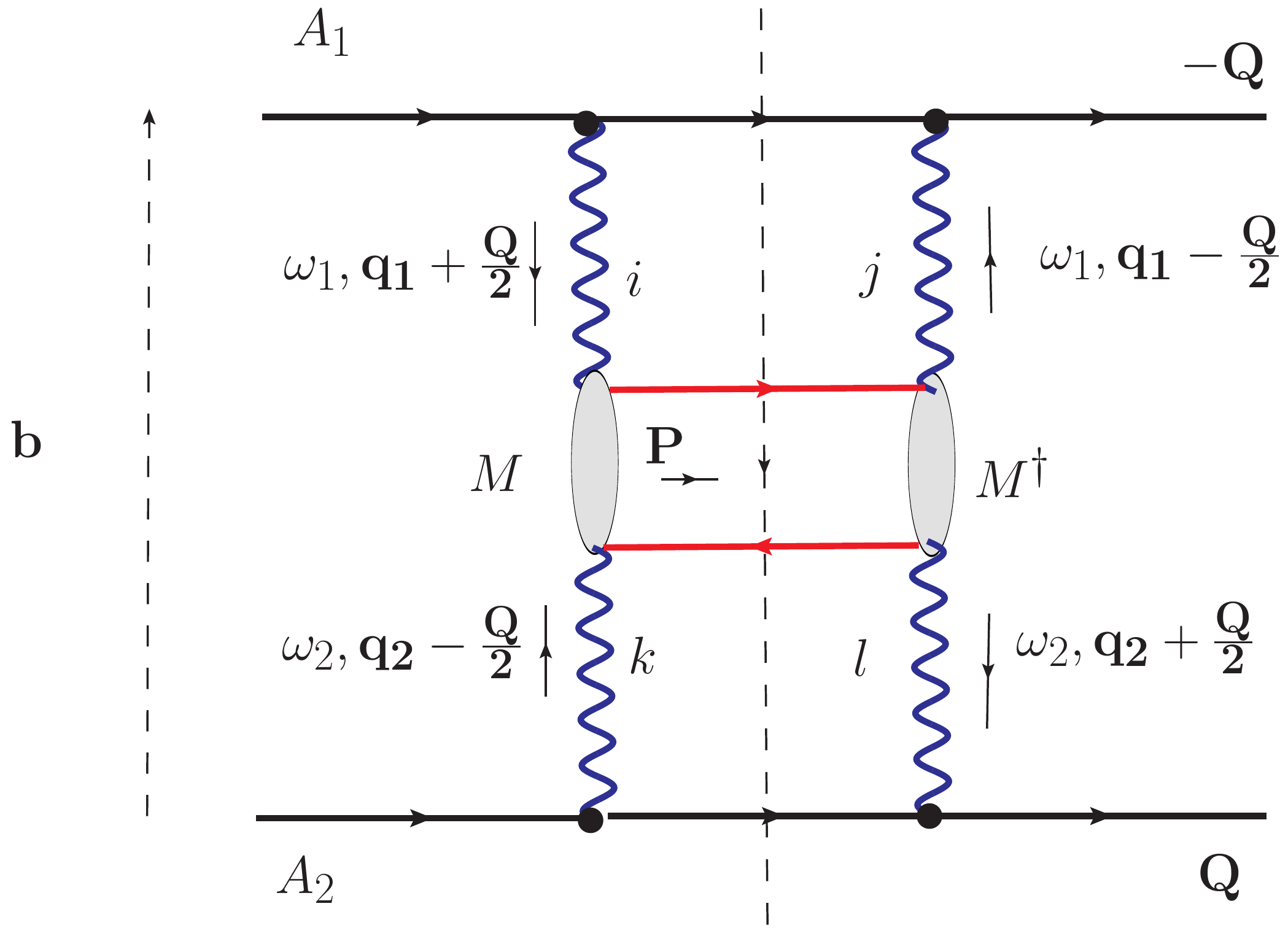}
	\caption{The cut off-forward $A_1 A_2 \to A_1 A_2$ amplitude. Its Fourier transform w.r.t. $\bQ$ yields the impact-parameter dependent cross section of dilepton production.}
	\label{fig:Feynman}
\end{figure}
The new distribution of photon implementation proposal concerns the use of a Wigner function. Then the standard photon fluxes in the momentum space and the impact parameter space, are obtained after integration over impact parameter or momentum space, respectively. 
The phase space of the dileptons can be parametrized
by the rapidities $y_{1,2}$ and transverse momenta
$\bp_{T,1}, \bp_{T,2}$ of leptons.
Then the cross section fully differential in lepton variables is obtained as
\begin{eqnarray}
{d \sigma [{\cal C}]\over dy_1 dy_2 d^2\bp_{T,1} d^2\bp_{T,2}} &=& 
\int {d^2 \bQ \over 2 \pi}  w(Q; b_{\rm max}, b_{\rm min}) 
\int {d^2\bq_1 \over \pi} {d^2\bq_2 \over \pi} \, \delta^{(2)}(\bp_{T,1} + \bp_{T,2}-\bq_1 - \bq_2)
\nonumber \\
&\times&   E_i \Big(\omega_1, \bq_1 + {\bQ \over 2} \Big) E^*_j \Big(\omega_1, \bq_1 - {\bQ \over 2} \Big)
E_k \Big(\omega_2, \bq_2 - {\bQ \over 2} \Big) E^*_l \Big(\omega_2, \bq_2 + {\bQ \over 2} \Big) \nonumber \\
&\times& 
{1 \over 16 \pi^2 \hat s^2}  \sum_{\lambda \bar \lambda} M^{\lambda \bar \lambda}_{ik} M^{\lambda \bar \lambda \dagger}_{jl} \, .
\end{eqnarray}
We perform the multidimensional integration using the VEGAS Monte Carlo
method \cite{Lepage:1977sw}.
The centrality of collision is determined by the values of the impact parameter, $(b_{\rm min}, b_{\rm max})$. We use an optical Glauber model \cite{Miller:2007ri} to estimate geometric quantities. The electric field depends on the charge form factor. In our ''Wigner function approach'' we use a charge form factor of the nucleus obtained from a realistic charge distribution described in \cite{KlusekGawenda:2010kx}. The cut off-forward 
$A_1 A_2 \to A_1 A_2$ amplitude is shown in Fig.\ref{fig:Feynman}.
Details of helicity structure of the $\gamma\gamma \to l^+ l^-$ amplitude and a lot more details about the research methodology can be found in the Ref. \cite{Klusek-Gawenda:2018zfz,ours2}.

\section{Results}

\begin{figure}[!h]
	\centering
	(a)\includegraphics[scale=0.35]{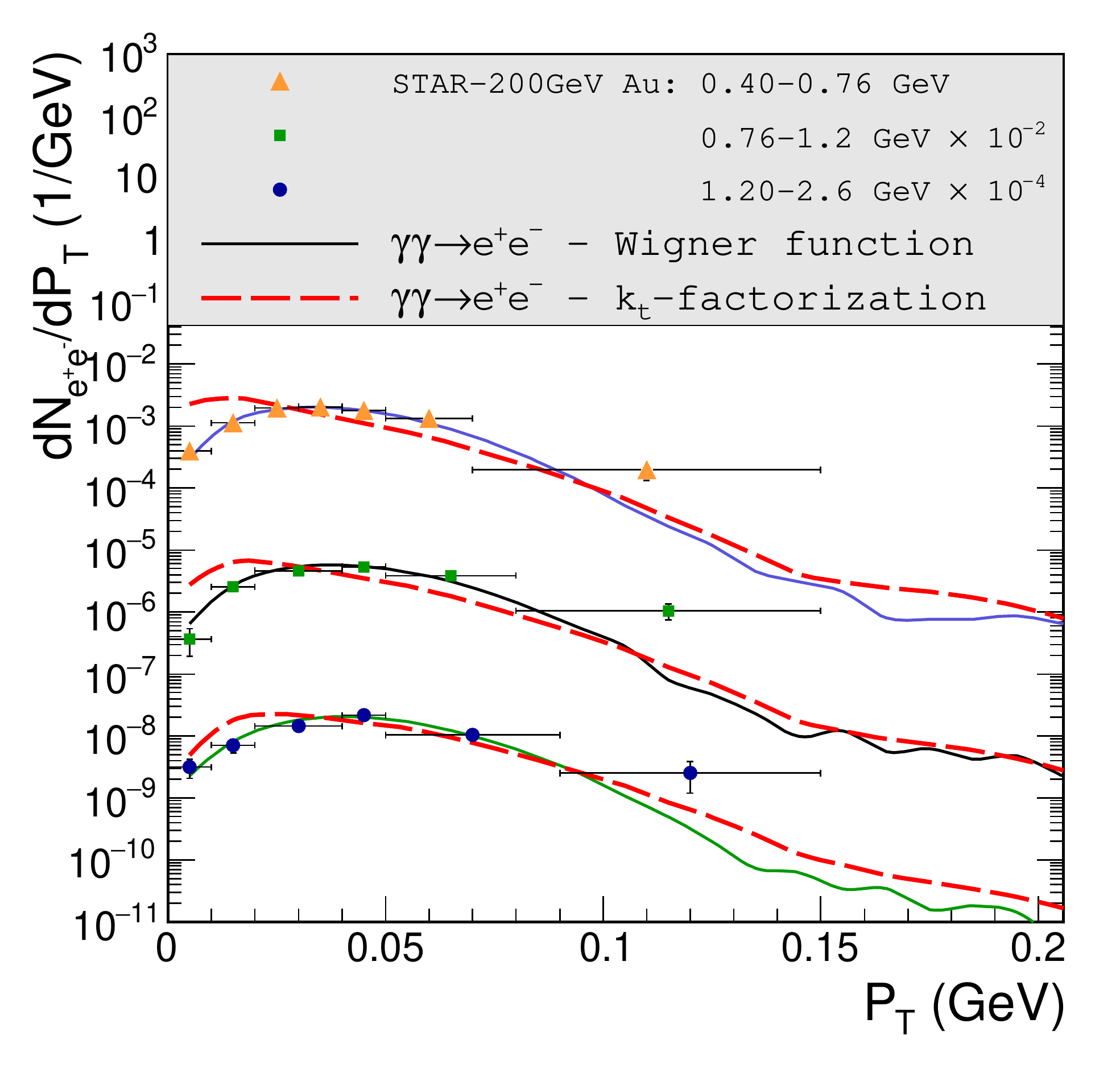}
	(b)\includegraphics[scale=0.35]{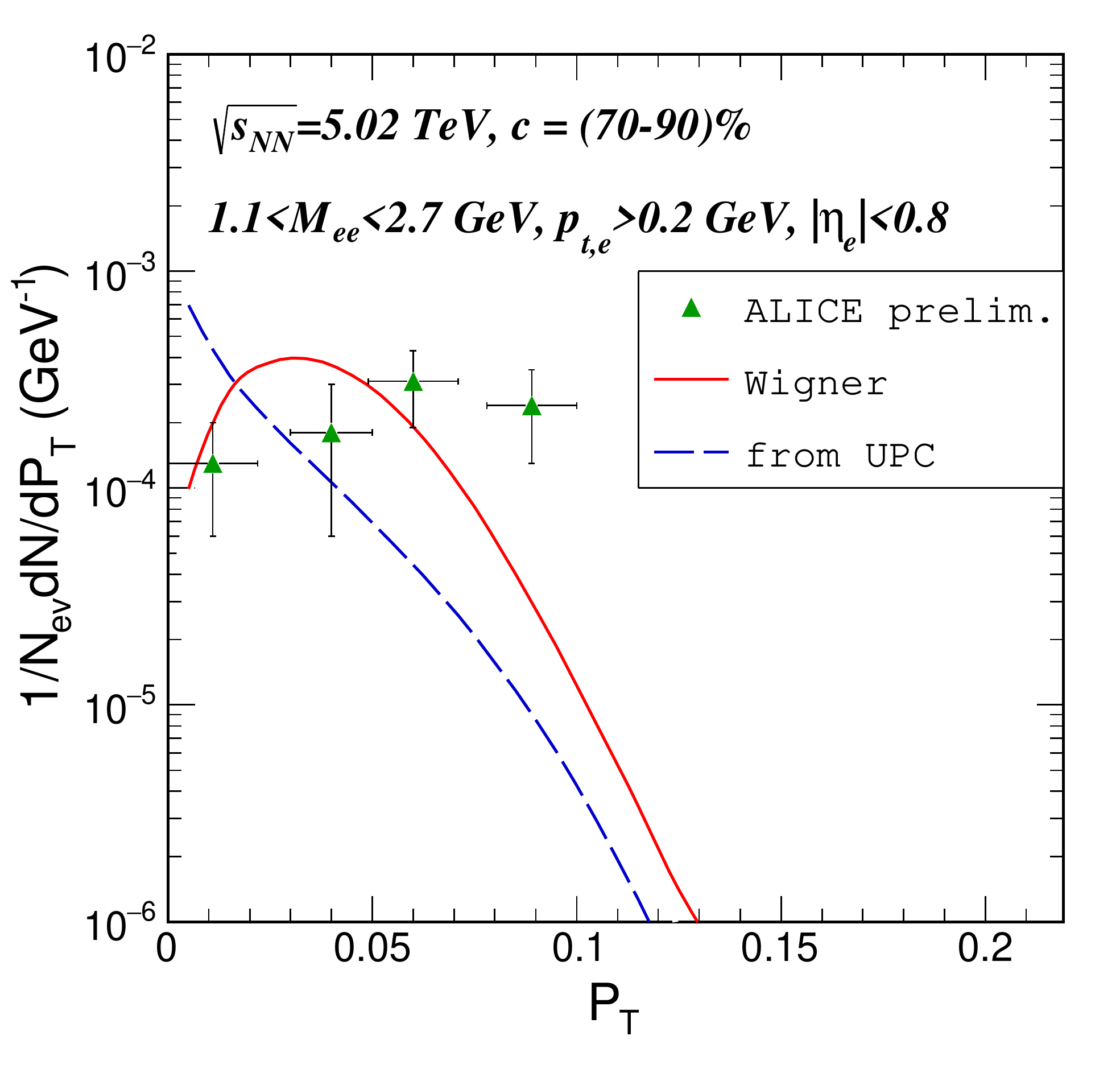}
	\caption{
		Distribution in transverse momentum of dielectron pair. Theoretical predictions versus STAR (a) and ALICE preliminary experimental data (b) are shown. We also include the experimental acceptance cuts on the single-lepton tracks as applied by STAR and ALICE detector. }
	\label{fig:ptpair}
\end{figure}

In Fig.~\ref{fig:ptpair} we show the distribution in transverse
momentum of the dielectron pairs. The STAR experimental data at RHIC energy ($\sqrt{s_{NN}}=200$~GeV) are summarized for three ranges of invariant mass and the centrality intervals: $(60-80)\%$. We get the excellent agreement of the newly presented approach (solid line - Wigner function) with the STAR experimental data.
In our previous work \cite{Klusek-Gawenda:2018zfz} we calculated the
$P_T$-distribution in what we dub a $k_T$-factorization approach,
where the transverse momentum of dileptons involves a
convolution of the $\bq$-dependent photon fluxes, schematically:
\begin{eqnarray}
N_{l^+ l^-}(\bP) \propto \int {d \omega_1 \over \omega_1} {d \omega_2 \over \omega_2} \int {d^2 \bq_1 \over \pi} {d^2 \bq_2 \over \pi}
\delta^{(2)}(\bP - \bq_1 -\bq_2) N(\omega_1, \bq_1) N(\omega_2, \bq_2) \sigma_{ \gamma \gamma \to l^+ l^-}(4 \omega_1 \omega_2) \, . \nonumber \\
\label{eq:kt-fact}
\end{eqnarray}
While our previous calculations in
\cite{Klusek-Gawenda:2018zfz} convincingly demonstrated the dominance of the $\gamma \gamma$ process at low $P_T$, the peak of the distribution obtained from Eq.~(\ref{eq:kt-fact}) is systematically at too low values of $P_T$. 
The left-hand side panel presents our results compared to the preliminary experimental results obtained by the ALICE collaboration \cite{Lehner:2019amb}. 
We get a similarly good description of the preliminary ALICE data as for the case of the STAR data.
For illustration, we show also the result of the $k_T$-factorization (dashed line)
proposed in \cite{Klusek-Gawenda:2018zfz}, which clearly fails to describe the position of the peak. 
Indeed, the peak of the $P_T$ distribution predicted by the $k_T$-factorization formula Eq.~(\ref{eq:kt-fact}) runs away towards smaller and smaller $P_T$ with increasing energy.
This is related to the fact, that in the photon distribution
\begin{eqnarray}
N(\omega,\bq) \propto |\bE (\omega,\bq)|^2 \propto {\bq^2 \over [\bq^2 + (\omega/\gamma)^2 ]^2} F^2_{\rm{ch}}(\bq^2 + (\omega/\gamma)^2 ) \, ,
\end{eqnarray}
the ``cutoff'' $\omega/\gamma$ decreases with increasing cm-energy (or boost $\gamma$ of the ion).

We have also studied low-$P_T$ dilepton production in ultrarelativistic heavy-ion collisions \cite{Klusek-Gawenda:2018zfz}, by systematic comparisons of the two sources that are believed to be prevalent in this regime, \ie, thermal radiation and photon-photon fusion within the coherent fields of the incoming nuclei. The former was taken from a well-tested model including in-medium hadronic and QGP emission rates, while the latter was calculated utilizing photon fluxes with realistic nuclear form factors including the case of nuclear overlap.
 
In Ref.~\cite{ours2}, we have shown the results for nine different centrality intervals, from central up to peripheral heavy-ion collision. There accoplanarity distributions for $\mu^+\mu^-$ production for the kinematics of the ATLAS experiment 
\cite{Aaboud:2018eph, ATLAS:2019vxg} were studied. Finally, we have a very nice description of the data including correct normalization and shape of the distributions. At larger acoplanarity, $\alpha = 1-{|\Delta \phi_{l^+ l^-}| \over \pi}$, there is a long tail possibly due to soft photon emission \cite{Klein:2020jom}.

In UPCs the incoming nuclei do not touch, \ie, no strong interactions occur between them. In this case one usually imposes the constraint $b > 2 R_A$ when integrating over
impact parameter $b = |\bb|$.
Here we left this restriction allowing the nuclei to collide. Then the final state will no longer contain the intact nuclei but the dileptons will be produced on top of the hadronic nuclear event characterized by an impact parameter $\bb$ (or range thereof). It seems to be interesting to check where, in the impact parameter space, dilepton production is the most prominent. Inside or outside the nucleus? From these, we have calculated, for the first time, the values of cross-sections for different regions in the impact parameter space. We take into consideration four ranges, as follows:
\begin{enumerate}
	\item full impact parameter space;
	\item outside of both nuclei: $b_{1}>R_{1}, b_{2}>R_{2}$;
	\item inside nucleus excluding the area of overlap: $b_{1}<R_{1}, b_{2}>R_{2}$ or $b_{1}>R_{1}, b_{2}<R_{2}$;
	\item overlapping area: $b_{1}<R_{1}$, $b_{2}<R_{2}$.
\end{enumerate}
In this case, we make very detailed calculations. We take into account the fact that the nuclear radius is not clearly defined and depends on the $R_0$ parameter:
\begin{equation}
R = R_0A^{1/3}
\label{eq:R}
\end{equation}
where $A$ is the atomic mass number (for Pb, A = 208). We have considered the range $R_0=(1.1 - 1.25)$~fm, which give $R\approx(6.52 - 7.41)$~fm for $^{208}$Pb.

\begin{figure}[!h]
	\centering
	(a) \includegraphics[scale=0.345]{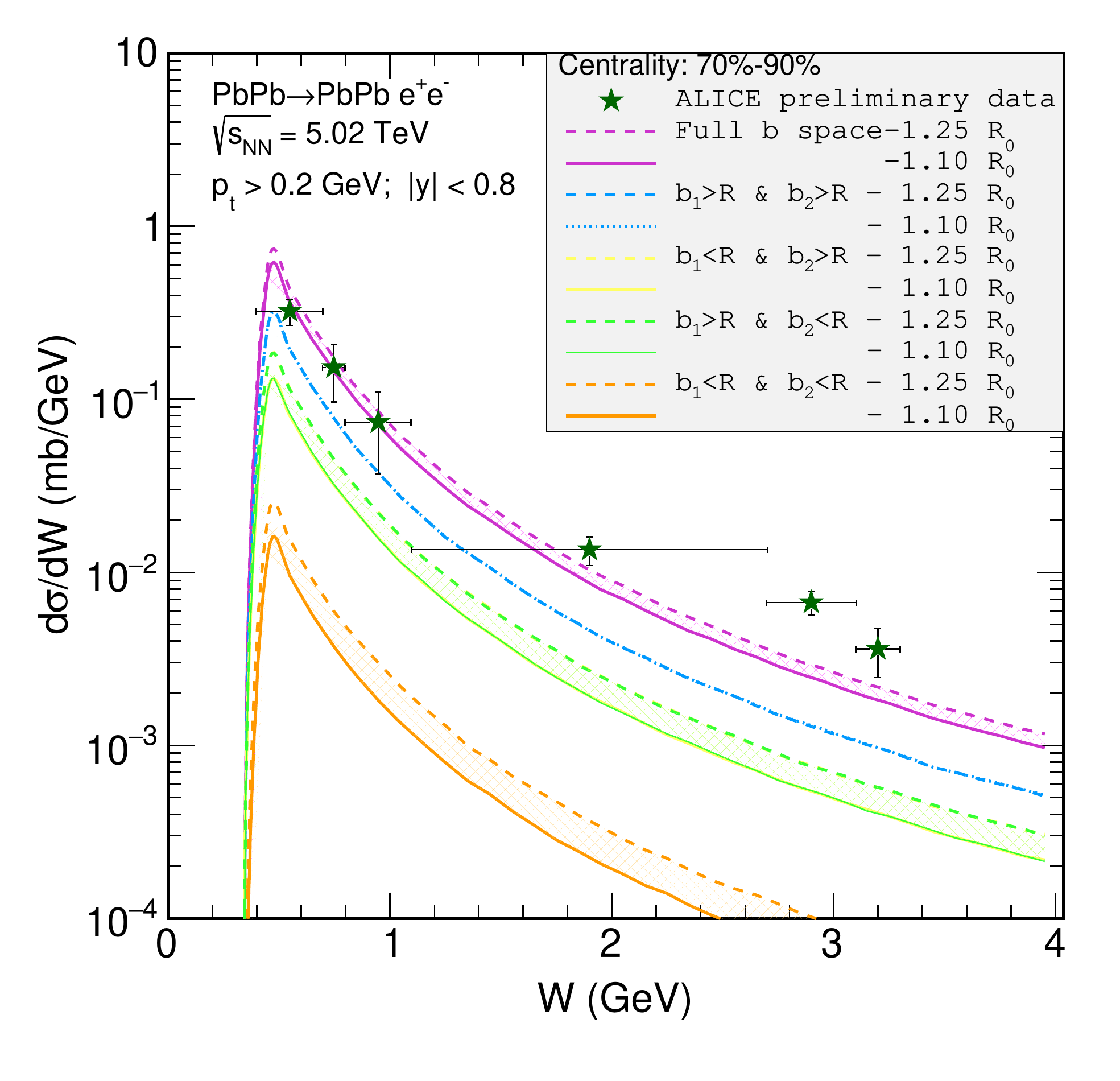} 
	(b) \includegraphics[scale=0.345]{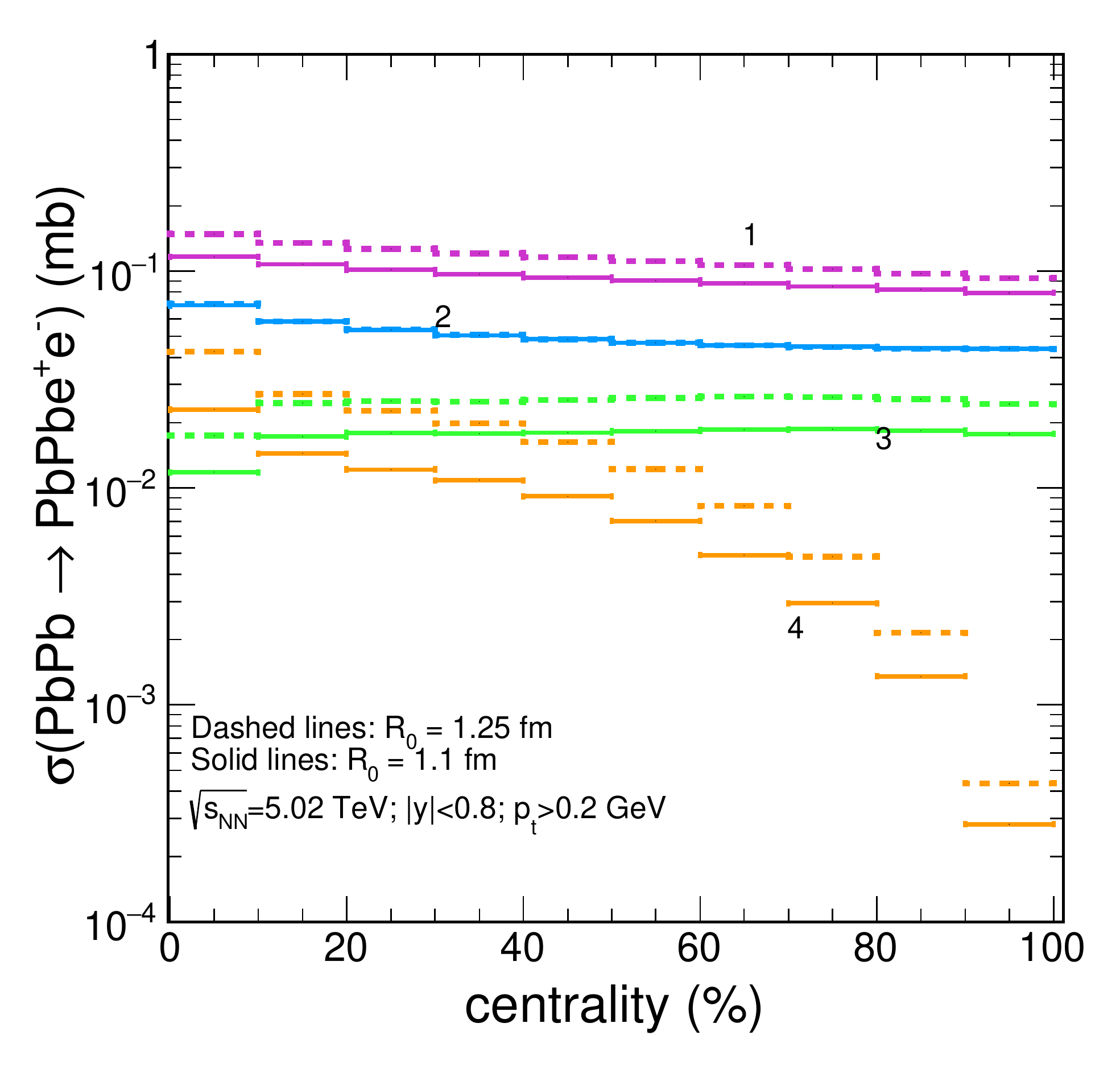}
	\caption{(a) Differential cross sections as a function of $e^+e^-$ invariant mass, $W$, for $p_{T,e^+e^-}<0.1$~GeV in $(70-90)\%$ central Pb-Pb collisions compared to ALICE preliminary data \cite{ALICE prem}. (b) Total cross section (mb) for PbPb $\rightarrow$ PbPb e$^+$e$^-$ process relative to collision centrality. 1-4 indication  corresponds to the four cases of the impact parameter space discussed in the text. In both graphs, dashed lines refer to $R_0=1.25$ fm and solid lines to $R_0=1.1$ fm in definition of nuclear radius (Eq.~\ref{eq:R}).}
	\label{fig:ALICE_ips}
\end{figure}

In Fig. \ref{fig:ALICE_ips} we present our results for the invariant mass distribution of low-$P_T$ dileptons. Theory is correlated with ALICE preliminary data \cite{ALICE prem} for the peripheral heavy-ion collision, $c=(70-90)\%$. The kinematical cuts correspond to ALICE di-lepton measurements at the collision energy $\sqrt{s_{NN}}=5.2$~TeV. We nicely describe experimental data, which indicates that the photon-photon process dominates for peripheral collisions. The left-hand side panel presents the relation between the value of the total nuclear cross section and collision centrality. In the case when the nuclei overlap each other (condition no. 4), with larger centrality, the common area is smaller, so the probability is drastically reduced. We have found that the ratio of the total cross section for the $e^+e^-$ production inside (condition 3+4) and outside of the nuclei gives $\approx$ 1. The cross sections for conditions closer to inside the nuclei show that the size of $R$ gets more and more significant (i.e., the differences in cross-sections become bigger).

\section{Conclusion}

We have presented results for centrality dependence of dilepton production focusing on $\gamma\gamma$ process. The implementation of Wigner distribution of photons in nuclei allows making calculations in the transverse momentum space, including centrality range. Our results, which have no free parameter, show good agreement with the STAR, ALICE and ATLAS experimental data.
The interplay of thermal radiation with the initial photon annihilation process triggered by the coherent electromagnetic fields of the incoming nuclei, studied in Ref. \cite{Klusek-Gawenda:2018zfz} was presented.
We have verified that the combination of photon fusion, thermal radiation, and final-state hadron decays gives a fair description of the low-$P_T$ dilepton mass spectra and dilepton transverse momentum distribution as measured by the STAR collaboration for different centrality classes, including experimental acceptance cuts. 
Recently the CMS collaboration has measured modification of $\alpha$ distributions correlated with neutron multiplicity \cite{Sirunyan:2020vvm}. A very new ATLAS study also presents the dimuon cross section in the presence of forward and/or backward neutron production \cite{Aad:2020dur}. So, since we have a tool that allows us to determine distributions in midrapidity and to control a number of emitted neutrons \cite{neutron}, we plan to study it in the future.

\end{document}